TMT.AOS.JOU.14.007.REL01

# NFIRAOS First Facility AO System for the Thirty Meter Telescope

Glen Herriot[*a], David Andersen[a], Jenny Atwood[a], Corinne Boyer[b], Peter Byrnes[a], Kris Caputa[a], Brent Ellerbroek[b], Luc Gilles[b], Alexis Hill[a], Zoran Ljusic[a], John Pazder[a], Matthias Rosensteiner[a], Malcolm Smith[a], Paolo Spano[a], Kei Szeto[a], Jean-Pierre Véran[a], Ivan Wevers[a], Lianqi Wang[b], Robert Wooff[a]

[a]National Research Council Canada, 5071 West Saanich Rd, Victoria, BC, Canada V9E 2E7;
[b]Thirty Meter Telescope, 1111 South Arroyo Parkway, Pasadena, CA USA 91105

## ABSTRACT

NFIRAOS, the Thirty Meter Telescope's first adaptive optics system is an order 60x60 Multi-Conjugate AO system with two deformable mirrors. Although most observing will use 6 laser guide stars, it also has an NGS-only mode. Uniquely, NFIRAOS is cooled to -30 °C to reduce thermal background. NFIRAOS delivers a 2-arcminute beam to three client instruments, and relies on up to three IR WFSs in each instrument. We present recent work including: robust automated acquisition on these IR WFSs; trade-off studies for a common-size of deformable mirror; real-time computing architectures; simplified designs for high-order NGS-mode wavefront sensing; modest upgrade concepts for high-contrast imaging.

## 1. INTRODUCTION

NRC-Herzberg in Victoria, Canada is designing NFIRAOS, which will be the first-light Adaptive Optics System for the Thirty Meter Telescope[1,2]. NFIRAOS is a Laser Guide Star (LGS), Multi-conjugate Adaptive Optics System (MCAO) system that will provide atmospheric turbulence correction in the near infrared over a 2 arcminute field of view and feed up to 3 instruments. In this paper, we present the design of NFIRAOS and then describe recent work and trade-off studies to reduce risk and to advance manufacturability, observing performance and efficiency on-sky.

## 2. DESCRIPTION OF NFIRAOS

### 2.1 Overview of NFIRAOS

NFIRAOS is supported on a Nasmyth platform of TMT and will provide near-diffraction-limited performance over the central 10-30 arcsecond field of view. NFIRAOS includes six laser guide star wavefront sensors, one high-order natural guide star (NGS) wavefront sensor for observations without laser guide stars, and one truth wavefront sensor (TWFS). There are two deformable mirrors (DM), one of which is mounted on a tip/tilt stage (TTS). NFIRAOS contains a source simulator (for natural objects and laser beacons), a phase screen and all associated entrance windows, beamsplitters, fore-optics, opto-mechanical devices, cooling, electronics and computing systems. It also includes test equipment, which is composed of a high-resolution wavefront sensor plus acquisition camera and miscellaneous fixtures.

### 2.2 Performance Requirements of NFIRAOS

The top-level requirements for NFIRAOS are as follows:

- Throughput: 80%, from 0.8 to 2.5 mm
- Background thermal emission: < 15 % of sky and telescope
- Wavefront Error: 187 nm RMS on-axis, and 191 nm over a 17" field of view (FoV)

*Glen.Herriot@nrc-cnrc.gc.ca; phone 1 250 363-0073; fax 1 250 363-0045; www.nrc-cnrc.gc.ca/eng/rd/nsi

- Sky coverage: 50 % at the Galactic pole (probability of meeting wavefront requirement in median conditions)
- Differential photometry: 2% for a 2 minute exposure on a 30" FoV at $\lambda = 1\ \mu m$
- Differential Astrometry: 50 mas for a 100 s exposure on a 30" FoV in the H band, decreasing with square root of time to a systematic floor of <10 mas
- Available from standby: <10 minutes
- Acquire a new field: < 5 minutes
- Downtime: < 1 per cent loss of science time due to failures and unexpected repairs

## 2.3 Science Path Optics

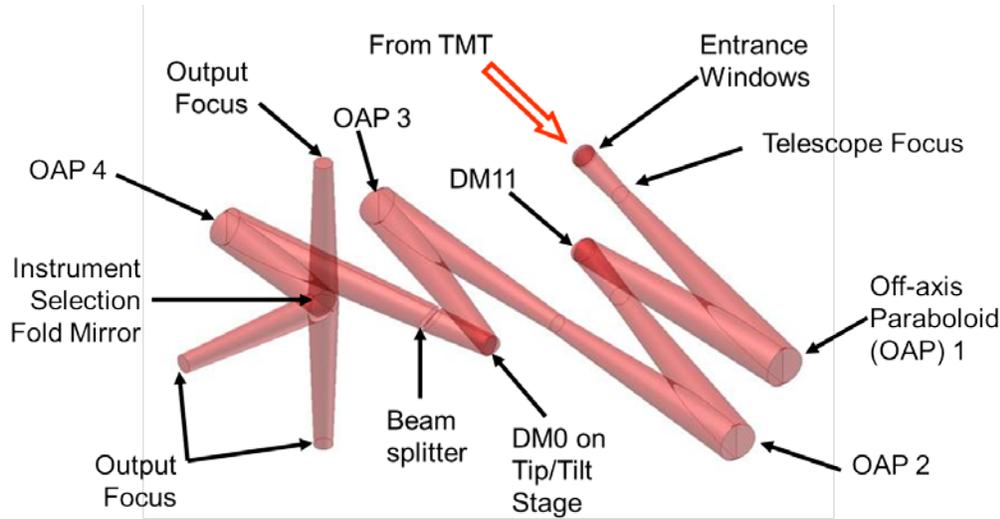

Figure 1 Science Path Optics

Figure 1 depicts the path that science light takes through NFIRAOS[12]. All of the optical elements shown lie in one horizontal plane. Light enters NFIRAOS through a pair of windows separated by a vacuum, depicted at the top of the figure. It passes through the focal plane from the telescope and continues to an OAP relay formed by two off-axis paraboloids[13] that collimate the beam onto a deformable mirror DM11, conjugated to 11.2 km. Both DMs have 5 mm pitch in the vertical direction and slightly longer in the horizontal direction. Following this relay, a second OAP pair contains a ground-conjugated deformable mirror DM0, mounted on a tip-tilt stage, and a beamsplitter. Finally, the fourth OAP4 reimages the beam and a steerable Instrument Selection Mirror sends the beam up, down or sideways to stations for three client instruments.

These science-path optics are shown from the same viewpoint, in context in Figure 2 as the red beam, together with the laser light in yellow and the visible NGS light in green. (All near-IR light longward of 800 nm continues to the science instrument). In this figure telescope light enters the double-paned evacuated entrance window and passes through focus. Near the focal plane, are two pieces of deployable calibration equipment: source simulators for NGS and LGS. The latter translates along the beam to emulate range-distance variations of the sodium layer. After reflection from DM11, there is a deployable turbulence screen, which translates vertically in the beam. Currently we are considering moving it closer to DM11 so that it will be in double-pass. This change would have the advantage of creating two turbulent layers closer in altitude to the real atmospheric turbulence. After OAP3 the tip/tilt stage carrying a DM at height h=0 is visible, followed by the science beamsplitter on a changer mechanism. Light shortward of 800 nm is reflected out of the science beam and is shown heading down and to the right. Immediately the 589-nm laser light is split off by the lower right-hand beamsplitter and is reimaged by OAP4-L. Finally, the laser light is directed onto the LGS trombone that compensates for the sodium layer distance. Individual fold mirrors pick off laser guide stars and send each into the cluster of LGS WFSs, visible to the right of the Tip/Tilt stage in this view.

Returning to consider the natural visible light that passes through the lower-right (natural vs laser) beamsplitter, we see that OAP4-V reimages this light, and on its way to focus, a pair of pointing and centring mirrors pick one natural guide star and send it to the VNW bench. The first of these mirrors is shown directly below the tip/tilt stage. The second is on the lower edge of the VNW bench, before a collimator and atmospheric dispersion corrector. For observing without lasers, this bench contains a 60x60 WFS for NGS SCAO mode. Because this WFS uses quad cells of pixels, there is a fast steering mirror to dither the spots on the 60x60 NGS WFS to estimate centroid gain as seeing changes. For LGS mode, a motorized fold mirror directs the light to a 12x12 Truth WFS to detect quasi-static aberrations. These especially include spherical aberration and Zernike 21 induced by aliasing from the sodium layer onto the LGS WFSs, plus non-common path calibration errors together with changes in telescope aberrations that vary with range distance to the laser guide star.

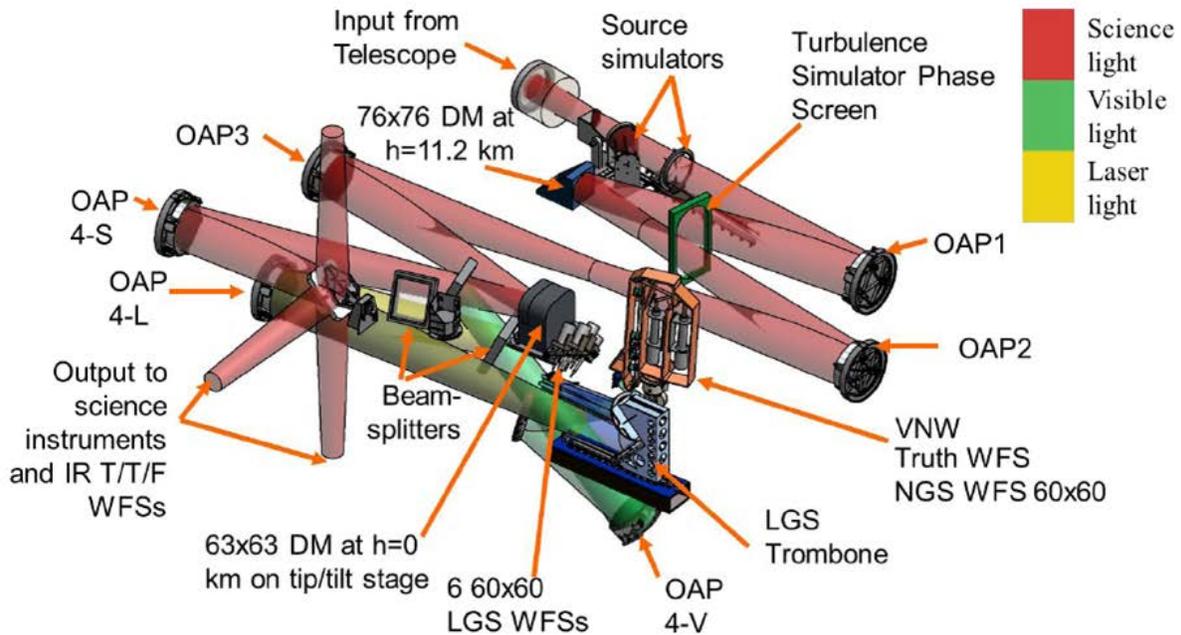

Figure 2 Isometric view of all NFIRAOS optics

All of the items in Figure 2 are contained in a cooled enclosure operated at -30 Celsius to reduce thermal background and reduce observing time. In K band, spectroscopic observations between the OH lines will be 2.5 times faster than if operating at dome-temperature on Mauna Kea.

### 3. TRADE-OFF STUDIES FOR A COMMON-SIZE OF DEFORMABLE MIRROR

During 2013, we undertook an extensive trade study on changing the size of one or both DMs, with the aim to reduce risk and deformable mirror cost. However, in the end, we chose *not* to redesign NFIRAOS. The two deformable mirrors in the baseline design of NFIRAOS are of different size: DM11 has 76 actuators across the diameter and DM0 has 63 actuators across its diameter. These are large and expensive mirrors, and a DM failure would cripple NFIRAOS. DM vendors indicated that if both mirrors were the same (preferably small) size it would reduce cost, risk and manufacturing time.

On the face of it, if DM0 were to break, then NFIRAOS would be inoperable, which would be an especially severe setback during integration and commissioning, because the team would be at a standstill. Full commissioning and even first light on TMT (defined as diffraction-limited images) would be delayed for several years until a replacement DM was obtained. Budgets and manufacturing capacity make it unlikely that we actually would have a spare DM on-hand during this critical phase.

Alternatively, if DM11 broke, in principle, a flat mirror could replace it, and NFIRAOS could operate in LTAO mode. Although LTAO falls far short of the scientific requirements for NFIRAOS, some astronomical observations are possible.

Another issue is that, in a sense, NFIRAOS is overdesigned. The baseline design of NFIRAOS has an identical 5-mm actuator pitch on both DMs. However, the optical design of NFIRAOS reproduces an identical beam-print of 300 mm-diameter for single objects on both DMs. This configuration results in the number of actuators in the metapupil at 11 km altitude being higher than needed to adequately correct the high altitude atmospheric turbulence measured in the TMT site survey data[9]. We found that a DM11 with fewer actuators could enable NFIRAOS to meet the requirements given in section 2.2, while saving money on actuators and drive electronics.

For all of the above reasons, we studied the possibility of redesigning NFIRAOS with two DMs of the same size. In that way, DM11 could serve as a spare for DM0. We considered having both DMs equal in size to DM0, or both equal to DM11 size, or to some intermediate size (68 actuators across diameter). We examined a series of variants that changed the size of the pupil and metapupil. The main consideration was to find an optical prescription that met all the existing requirements such distortion, exit pupil location, and overall space envelope, while incorporating the subassemblies described in section 2.

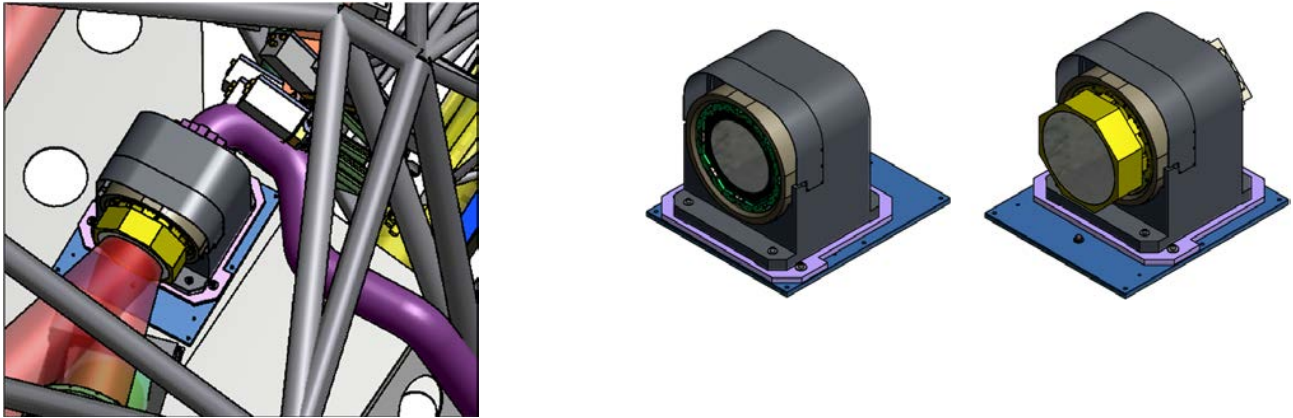

Figure 3 Tip Tilt Stage can carry DM11 in case of DM0 failure. The left figure shows the TTS in NFIRAOS in the fall-back position with the TTS moved back and DM11 extending in front of the stage. The middle image shows the baseline configuration with DM0 mounted on the TTS, and the right shows the repair position.

Initially we believed that if the ground-conjugated DM diameter were larger than the baseline DM0 of 63x63 actuators, then we would need a new and larger tip-tilt stage, because DM0 nests within the tip-tilt stage. Since we already have bought the tip-tilt stage and plan simply to upgrade its electronics before installing it in NFIRAOS, buying a new stage would be expensive. However, we later realized that DM11 could be flush-mounted on the tip-tilt stage. This attachment would move the optical surface forward, but to compensate, the stage itself could be moved back onto a second mounting location in NFIRAOS (see Figure 3). The stage would need additional counterweights to balance the larger DM11 whose centre of gravity would then lie ahead of the stage tilt axis. We feared that the additional moment of inertia would be beyond the capability of the existing voice-coil actuators to provide adequate bandwidth with acceptable heat dissipation.

However, as shown in Figure 4 we modeled the effect on performance and were pleasantly surprised. First, we assessed the tip-tilt performance with the baseline DM0 on the TTS and then changed the DM moment of inertia to match DM11, retuned the controller and evaluated performance. It turns out the increased moment of inertia of DM11 will reduce the TTS bandwidth from about 80 Hz to 50 Hz.

This bandwidth is well above the 20 Hz woofer-tweeter split, i.e. high temporal frequency tip-tilt is corrected by the DM0 surface itself, and only low frequencies by the TTS. Thus, the TTS does not see commands above 20 Hz, and we do not expect a performance reduction for tip/tilt control.

As we tried to package the optical variants, we recognized that increasing either DM diameter was expensive, and that the opto-mechanical effort to shrink DM11 and/or expand DM0 was running into space constraints. The follow-on effort and risk to rework the structure, subassemblies and thermal enclosure for NFIRAOS posed a large risk to our schedule and resources. Taken together with the realization that DM11 can act as a temporary spare for DM0, we decided to retain the baseline design with DM0 having 63x63 actuators and DM11 having 76x76 actuators.

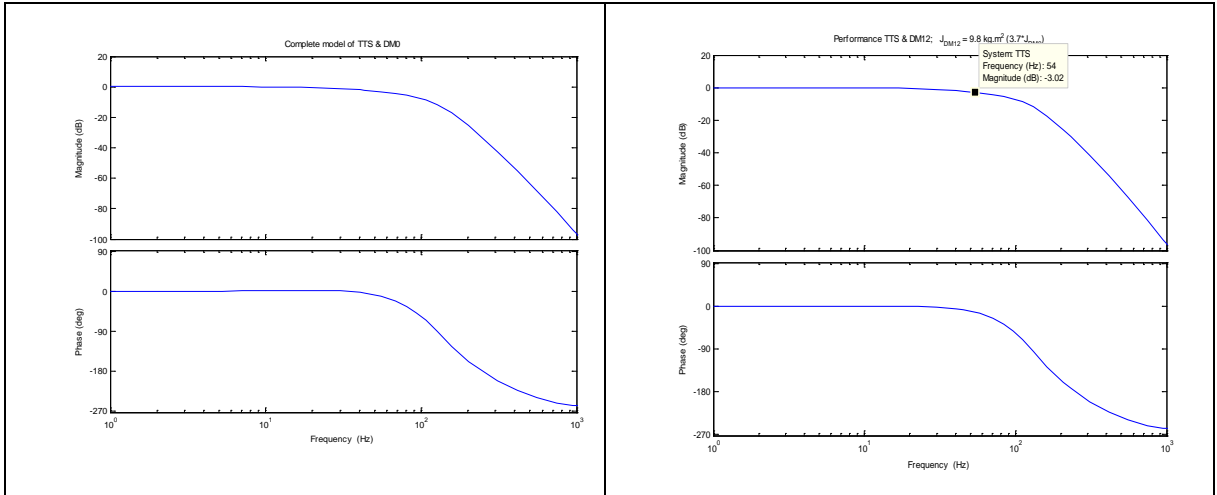

Figure 4 Bode Plots of tip-tilt stage performance with DM0 (left) and DM11 (right)

## 4. REAL-TIME COMPUTING ARCHITECTURE TRADE-OFF STUDY

As well, during 2013, we revisited computing architectures[4] for the Real Time Controller in NFIRAOS. At the time of our last major NFIRAOS review in late 2011, the baseline for the RTC used boards with 6 Field Programmable Gate Arrays running an iterative algorithm. Because the RTC ingests about 35 K slopes from illuminated WFS subapertures and controls nearly 8K DM actuators, the number of arithmetic operations and memory accesses per 800 Hz cycle is nearly one hundred times larger than on existing AO systems. The iterative algorithm technique drastically reduced the number of computations compared with a standard Matrix-Vector Multiplication (MVM), but was complex to explain to programmers and difficult to program. The planned FPGA boards were developed for radio-astronomy, but with adaptive optics requirements also in mind. The custom nature of these FPGA boards and the labour-intensive specialist programming required meant that this approach was costly and acknowledged as difficult to develop and maintain.

Meanwhile advances in commercial computation offered hope that an RTC was within reach using more conventional techniques and available boards. We benchmarked MVM on clusters of CPU servers[5], with and without accelerators both Graphical Processing Units (GPUs)[6], and Intel Xeon Phi. Half a dozen dual-GPU boards can do the required computation in ~ 900 $\mu$s with jitter of a few 10s of $\mu$s, which gives good margin compared with the frame period of 1.25 ms. While we found that a cluster with 12 Xeon Phis computes most frames quickly enough, occasionally the internal operating system suspends the RTC work, resulting in unacceptable jitter of up to 10 milliseconds in the worst case. Since hard real-time control is not the marketing niche for the Xeon Phi, and its operating system is only available from Intel with no real-time upgrade expected, we have ruled it out for our RTC.

Using CPUs alone with a real-time patch applied to the Linux operating system, our benchmarks showed that six servers, each with two dual-CPU motherboards can process the WFS data and create DM command vectors on average <800 microseconds, with low jitter resulting in a worst case of slightly less than 900 microseconds as shown in Figure 5. For this benchmark, simulated pixels for 1/2 LGS WFS were streamed in a 500-$\mu$s burst over 10 Gb/s Ethernet to one server which processed pixels (computing slopes and statistics) and then applied MVM to calculate its portion of the DM command vector and finally returned the entire vector over Ethernet to the pixel simulating computer. Note that this figure does not include the effect of swapping matrices periodically every 10 seconds. Swapping matrices adds 120 $\mu$s to those few frames – an acceptable latency.

In the all-CPU architecture, a dedicated non-blocking Ethernet switch sits at the heart of the RTC and connects all the components and routes all I/O to and from WFSs, DMs, compute engines, and telemetry storage. There are a total of 8 server chassis, each with two dual-CPU motherboards. Each of six chassis receives pixel data from an individual LGS WFS. A seventh chassis is the central housekeeper that handles the OIWFSs, coalesces the DM command vector and sends it via the switch to the DM electronics[10]. Finally, for reliability, an eighth server chassis is an on-line spare. In the event of a server failure, the high-reliability switch will be remotely reconfigured to replace the failed machine so that observing can continue that night with only a few minutes lost time.

Our baseline is now a pure-CPU architecture. Although GPU accelerators can reduce the number of server chassis and total power consumption, we concluded that ease of development, and minimizing the variety of hardware and software components in the RTC, made an all-CPU solution attractive, especially when considering maintenance over the lifetime of NFIRAOS. The Intel roadmap indicates that it may be possible to process an entire LGS WFS on only two CPUs as early as 2015, with the arrival of the E5-2600 v4 CPUs (we are currently using v2, but v3 is expected in this fall of 2014).

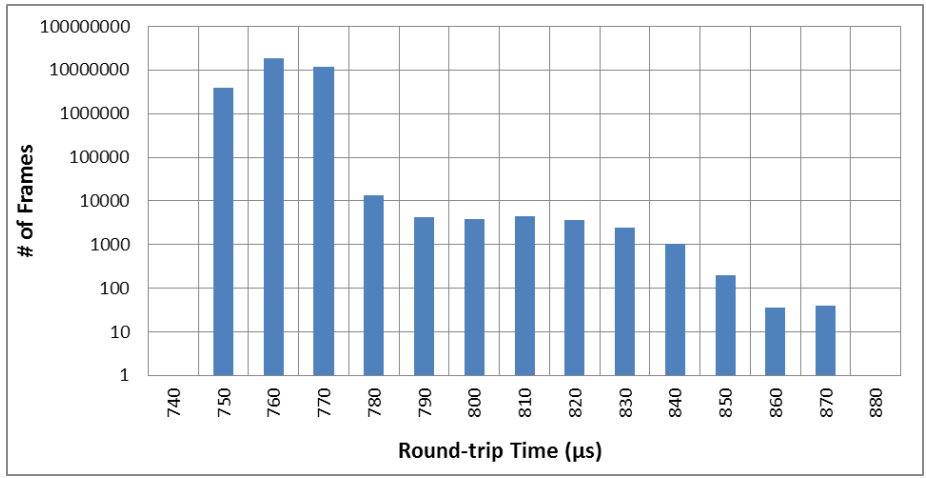

Figure 5  Benchmark timing of CPU-based RTC

## 5. ROBUST AUTOMATED GUIDE STAR ACQUISITION

Observing efficiency is an extremely important requirement of NFIRAOS. For high sky coverage, NFIRAOS must lock on tip/tilt/focus stars as faint as 20.5 magnitude in J band, and tip/tilt stars of up to $22^{nd}$ magnitude, shown in Figure 6. We have developed and simulated robust automated techniques to acquire and lock on natural guide stars quickly. This section of the paper explains the acquisition issues and our solution to them, together with the expected performance.

NFIRAOS relies on each client instrument having up to three near-infrared OIWFSs[8] (on-instrument WFSs) to control tip tilt and focus as well as rotation and plate-scale modes like magnification that are not well sensed by laser guide stars. These OIWFSs are individually configurable to be either an imager (tip/tilt sensor) or a 2x2 Shack-Hartmann WFS. For every observation, one will be a 2x2 WFS (sensing tip/tilt/focus on typically the brightest of the three NGS guide stars). The OIWFSs each use a 1k x 1k quadrant of a Hawaii 2RG near-infrared detector This large area helps in acquiring guide stars, and permits precise "electronic dithering" of the telescope without using guide-probe mechanical motion as described below in section 5.1. The optics create images that are detected with 5.6 or 11.2 milli-arcsecond pixels in T/T/F or 2x2 mode respectively, to Nyquist-sample diffraction-limited images composed of H+K band photons. NFIRAOS relies on the LGS measurements and two deformable mirrors to correct the high order aberrations of the natural guide stars and thus sharpen these guide stars.

However, to read out the entire 1k x 1k area takes about 0.7 second. At that rate, during a science exposure, the OIWFS readout speed and noise performance would be insufficient to achieve the NFIRAOS sky coverage and wavefront error requirements. Therefore, in operation, with all control loops closed, we plan to read out only small 4x4 pixel regions of interest (ROI) employing multiple co-adds on the detector. For median brightness stars (17.5 magnitude) our expected

OIWFS frame rate is 90 Hz, but for the brightest guide stars, the rate will go up to match the 800 Hz rate of the LGS WFSs. On the OIWFS configured as a 2x2 WFS for sensing tip/tilt/focus, there are 4 ROI windows, while the other two OIWFSs that sense tip/tilt have one ROI window each on their detectors. These ROIs are centred on the 1k imager (or quadrant of the imager) and are embedded in larger 10x10 guard windows digitized less frequently, with fewer co-adds, resulting in worse noise for these pixels. This scheme optimizes frame rate and noise while providing insurance against transients causing loss of lock.

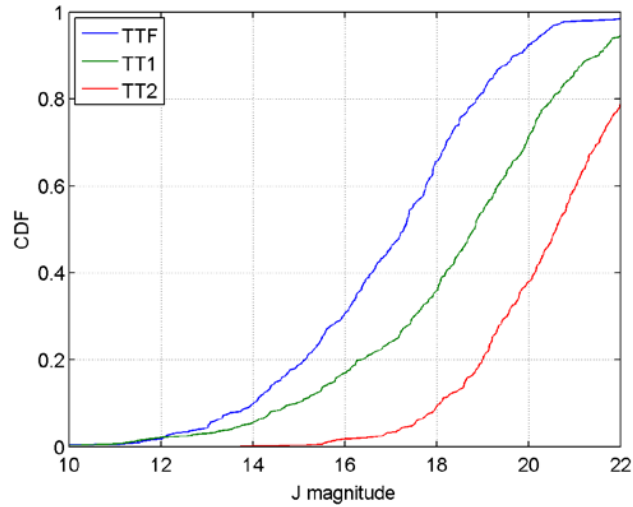

Figure 6  Cumulative probability of guide star magnitude

Another issue is that the blind pointing requirement of TMT is 1 arcsecond RMS on the sky, and the plan for TMT is to start night-time commissioning of NFIRAOS soon after installing the last primary mirror segment, while telescope pointing models will likely still being refined. As a result, after slewing the telescope to a new field of view, the guide stars will probably not be well-centred on the OIWFSs, and there is reasonable change that a guide star may not even fall anywhere on the 5.6 arcsecond field of view of the entire 1k x 1k detector. Consequently, at first light, NFIRAOS's 20-arcsecond square acquisition camera will occupy the side instrument port. Eventually with three instruments, there will always be a science imager to serve as an acquisition camera for itself or its neighbours. NFIRAOS's instrument selection mirror is required to switch the beam accurately from one port to another in less than 10 seconds. Servo analysis of our design of the selection mirror mechanism predicts that it will operate in 5 seconds.

During this 5-10 seconds, measurement errors on the acquisition image, telescope-tracking errors, together with windshake and atmospheric turbulence will cause the guide star image to be off-centre. These latter two effects alone are predicted to cause 90 mas RMS motion for $70^{th}$ percentile worst seeing and $95^{th}$ percentile windshake, so the image may be hundreds of pixels off-centre, and moving. Thus, initially we have to read a large ROI, which takes time and/or causes poor signal to noise. Furthermore, with longer delay times for photon integration and detector readout, servo loop gains must be reduced to preserve stability. Therefore, with low loop gains, poor image stabilization causes image trailing, further reducing signal to noise ratio and measurement accuracy. For all of the above reasons we have developed the process described in the next sections.

### 5.1 Acquisition Procedure

- Guide stars and science targets and field rotation angles are pre-planned before observing
- When acquisition begins, these actions are done in parallel:
    - The telescope slews to point the acquisition imager directly at the natural guide star for the 2x2 OIWFS
    - The three OIWFS probes are moved to their mechanical locations planned for the observation

- o The instrument-selection fold mirror in NFIRAOS directs the beam to the acquisition imager
- The high order AO loops are closed between the LGS WFSs and DMs to correct atmospheric turbulence, but not tip/tilt or plate scale
- An image is taken on the acquisition camera of the Tip/tilt/focus star (usually brightest of three guide stars)
- In parallel:
  - o The instrument selection fold mirror repositions the NFIRAOS output beam towards the science instrument
  - o A pointing correction offset is calculated from the acquisition image, and then the telescope slews to position the guide star on the centre of the 2x2 T/T/F OIWFS detector
- Tip-tilt and focus loop is gradually locked on the T/T/F guide star, with bandwidth increasing over several seconds as described in detail in section 5.2
- After stabilizing Tip/Tilt on one OIWFS, the remaining two guide stars will be close to their final positions on their detectors and with small residual tilt jitter.
- Finally, plate scale, rotation and tip/tilt loops are closed on these last two guide stars in parallel, by starting mid-way through the locking process for each.

## 5.2 Progressive locking onto guide star

The general principle is that as NFIRAOS locks on a guide star, the detector readout area is progressively reduced and the frame rate increased while servo parameters are adjusted. The entire process is deterministic and table-driven. The sequence of parameters for decreasing frame size is held in a table. The OIWFS camera electronics act on several columns of the table, and the RTC the remainder. Each row of the table describes one frame integrated and readout from the detector. During the locking, both the RTC and OIWFS step through the table synchronously. By the time the sequence is on the last table row, the guide star is locked on the 4x4 ROI, at a suitable frame rate. The system then repeats the last row endlessly during the science observation.

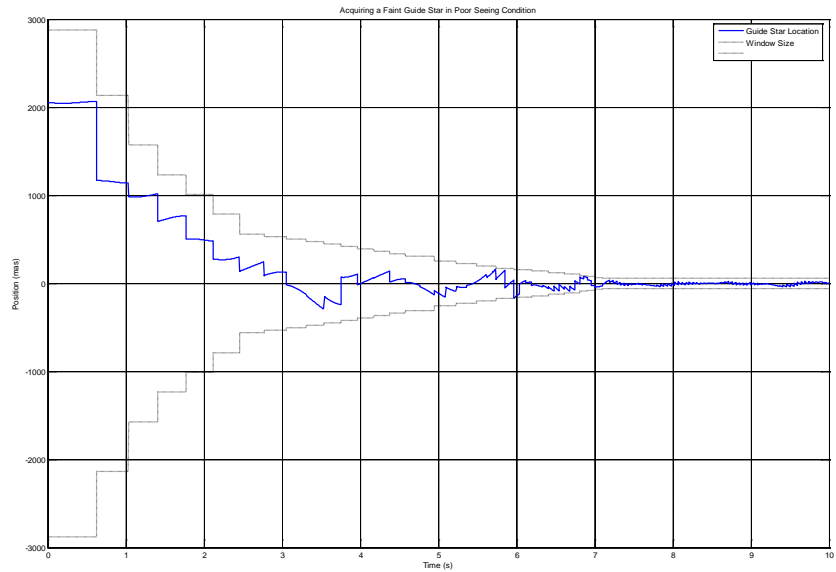

Figure 7  Acquisition of 20.5 magnitude guide star (worst case for T/T/F). X axis time (s). Y axis ROI (pixels)

In Figure 7 the dashed lines show the size of the ROI; steps in the dashed line correspond to when a new image is sent to the RTC. The blue solid line is the image location during integration on the OIWFS. Trailing is evident from the start. In the early rows of the sequence table, rapid acquisition is relatively more important than accuracy. Each pixel in the ROI is digitized once. However, as the ROIs become smaller, during the integration time, the camera electronics digitizes fewer pixels more often and co-adds them before sending the image to the RTC. It is always a tradeoff between readout time, read noise, photon noise, control loop stiffness and noise propagation. Because the tip/tilt stage and its control have finite bandwidth, we cannot be too aggressive in reducing the ROI from step to step, but must be patient to ensure that the image will have moved to a location within the next smaller window size. For example in Figure 7, starting at 3 seconds, there is a large disturbance at a time when we are only reading 4 frames per second. Because of this latency, loop gain is low and the correction takes some time to recover.

Our simulations show we can acquire a $14^{th}$ magnitude star in 1.5 seconds, and in 7 seconds a $20.5^{th}$ magnitude star, which is the faintest T/T/F star needed at the galactic pole. In Figure 7 we see the worst case, with the faintest star and $95^{th}$ percentile windshake and $70^{th}$ percentile seeing. Because the images are faint and trailed, we need relatively long integration times and a less-aggressive progression of servo gain and ROI reduction. Initially the control gain must be small because of the large latency, and so tip/tilt control has low closed-loop bandwidth. As the readout time decreases, the loop gain is increased, and more co-adds are used for each frame.

### 5.3 NGS Acquisition Centroiding

Ordinarily during observations, NFIRAOS measures the centroid positions of the natural guide stars using a matched filter algorithm[18]. Matched filters are noise optimal, and computed as a background task from a model image updated by seeing estimates from RTC telemetry. A matched filter can accurately measure sub-pixel image motion. However, such a matched filter has limited dynamic range of only a few pixels, and is unsuitable for locating a star throughout a large image, especially an image that has trailed during the initial long exposures. Therefore, the acquisition process begins using a brightest pixel algorithm. The early images in the sequence are median filtered and then brightest pixel identified. At a pre-determined row of the acquisition parameters table, when exposure times are sufficiently short and loop gains higher, producing compact images located in a smaller region of interest, then the algorithm automatically switches to using matched filters for centroiding.

## 6. SIMPLIFIED DESIGNS FOR HIGH-ORDER NGS-MODE WAVEFRONT SENSING

The preliminary design of NFIRAOS described in section 2, has two separate visible-light natural guide star WFSs inside the cooled enclosure. One is for controlling NFIRAOS in SCAO mode without lasers, (such as for high-contrast imaging) and the other WFS is a truth wavefront sensor. The baseline design of NFIRAOS, described in section 2, uses a pair of pointing and centring mirrors to select a natural guide star and direct it into a stationary visible natural guide star bench. On that bench a deployable fold mirror selects either NGS SCAO or LGS MCAO mode. In the former mode, light enters a 60x60 SH WFS with quad cells of pixels for each spot; for the LGS mode the light enters a separate truth WFS with 12x12 lenslets, and 8x8 pixels behind each. Two difficulties with this overall approach are that firstly the star selection mirrors require extremely precise angular control and one of them is almost 400 mm diameter; secondly two expensive low noise cameras are used.

Consequently, we have recently studied a compact design with a single camera. Flip mirrors divert light through separate lenslet arrays and reimage them onto one camera. This assembly is small enough that an XY stage may position it to select a single star. The result of these two changes should be better performance at lower cost.

## 7. MODEST UPGRADE CONCEPTS FOR HIGH-CONTRAST IMAGING.

The TMT astronomical community wishes to extend the capability of NFIRAOS and its client instrument IRIS[7] cost-effectively before a specialized exo-planet instrument can be built for TMT. Although not yet funded in the baseline NFIRAOS, we have studied a collection of some relatively simple upgrades that would greatly improve the capabilities of NFIRAOS for high-contrast imaging and spectroscopy of extra-solar planetary systems, purely using the NGS WFS to control NFIRAOS as an SCAO system. Future work will assist making a decision whether to proceed, and includes doing higher fidelity performance estimates and costing of the following proposed upgrades,

### 7.1 Potential high-contrast revisions to NFIRAOS

- Better polished optics in NFIRAOS, especially entrance windows

- Small-field K-mirror and relay optics deployable after the entrance windows
- Specialized beamsplitter with a shaped pupil and a grating superimposed
- Narrow-field ADC deployable in converging exit beam

None of NFIRAOS' optics are conjugate to either DM. The worst case is the entrance window that is relatively close to focus and 200 km from a pupil. As a result, the DMs cannot correct polishing errors on NFIRAOS' own optics, which results in quasi-static speckles that average out on time scales vastly longer than needed to average residual atmospheric turbulence. As an aside, polishing errors also cause astrometry[11] errors from high-spatial frequency image distortions, because beam prints on optics are different for objects scattered throughout the field of view. Therefore, to meet the astrometry requirement, we have recently tightened the polishing specification for the windows from 28 nm RMS to 3 nm.

Rotation of beam prints on optics during an exposure exacerbates the long speckle averaging time. Because TMT is an alt-az telescope, the image and pupil rotate, each at different rates. All client instruments remove image rotation by rotating themselves, and some like IRIS have a rotating pupil mask. But, because it is upstream from the instrument rotator, the beam prints on NFIRAOS optics rotates. To reduce this we have created a design concept for a deployable narrow-field K-mirror. This mirror fits behind the entrance windows, in front of focus and includes simple relay optics to preserve the location of the focal plane and telescope pupil as seen from NFIRAOS. With this K mirror, the beam is stabilized on all downstream optics, but not on the windows and the K-mirror itself.

The beamsplitter changer's second position, intended for an engineering beamsplitter, can be replaced with an optic that includes a beamsplitter, a shaped-pupil mask (e.g. an array of cat's eyes), and a regular grid of opaque spots or lines. The shaped pupil mask prevents the hard-edged pupil mask within IRIS from scattering light. The grid on the beamsplitter acts as a diffraction grating, causing satellite images in the focal plane of the instrument.

The final part of the upgrade package is a deployable small-field atmospheric dispersion compensator, located just inside the bottom exit port that feeds IRIS.

### 7.2 Companion high-contrast upgrades to IRIS

The deployable ADC and high-order high-bandwidth correction of turbulence by NFIRAOS means that it will deliver compact images to IRIS. Inside the IRIS imaging spectrograph, the plan is to have simple coronagraphic masks deployable from the tip of each OIWFS probe beyond the pickoff mirrors. These masks, one for each of J, H, or K bands, will block the star's light at the focal plane and keep it from the science instrument. At the same time, the diffracted satellite image of the star created by the beamsplitter grating described above will fall on the OIWFS pickoff mirror and serve for guiding. In this way, TMT can keep the starlight from the instrument but use the ghost image of the star for low-temporal frequency tip/tilt/focus correction.

## 8. CONCLUSIONS

We have advanced the design of NFIRAOS and studied risk reductions and simple upgrades to improve its scientific capabilities. This work has included trade-off studies for a common size of DM, RTC architectures and benchmarks, NGS and truth WFS optical and mechanical concepts. We have developed and simulated robust algorithms for rapid automated guide star acquisition. As well, we have begun to investigate affordable upgrades to improve high-contrast imaging for exoplanet spectroscopy. NFIRAOS is currently waiting funding approval from the Canadian government, and we hope to begin construction within a year.

## ACKNOWLEDGEMENTS

The authors gratefully acknowledge the support of the TMT collaborating institutions. They are the Association of Canadian Universities for Research in Astronomy (ACURA), the California Institute of Technology, the University of California, the National Astronomical Observatory of Japan, the National Astronomical Observatories of China and their consortium partners, and the Department of Science and Technology of India and their supported institutes. This work was supported as well by the Gordon and Betty Moore Foundation, the Canada Foundation for Innovation, the Ontario